# Strong Thermal Conductivity Suppression in Isotope Purified Single Crystal Diamond with Phonon Scatterings of Color Centers

Jinwen Liu[1,#], Junliang Fan[2,#], Zimu Jiang[1], Zifeng Huang[2], Zhe Cheng[1,2,*]

1 School of Software & Microelectronics, Peking University, Beijing 100871, China

2 School of Integrated Circuits and Beijing Advanced Innovation Center for Integrated Circuits, Peking University, Beijing 100871, China

[#]These authors contributed equally to this work.

*Author to whom correspondence should be addressed: zhe.cheng@pku.edu.cn

## Abstract

The thermal conductivity of natural diamond can be further improved by ~50% by eliminating isotope impurity to obtain pure $^{12}C$ diamond. However, unlike isolated point defects, how color centers scatter phonons and correspondingly limit the thermal conductivity remain poorly understood. Here, we report the thermal conductivities of two isotope-purified diamond with different Nitrogen-vacancy (NV) color center concentrations, confirmed by photoluminescence (PL) and secondary-ion mass spectrometry (SIMS). Results show ~ 3200 W $m^{-1}$ $K^{-1}$ and ~ 2100 W $m^{-1}$ $K^{-1}$ for two samples of low and high NV concentrations, respectively. Modified Callaway model analyses show that first-principles-derived scattering parameters can reproduce room-temperature thermal conductivity of the high NV concentration sample under a vacancy concentration assumption (~30% of the total Nitrogen concentration), but fail to reproduce its temperature dependence. Compared to PL and SIMS results, which show ~7.1-fold NV concentration and ~92-fold difference in total Nitrogen concentration between two samples, suggesting the assumed 30% vacancy fraction may be greatly overestimated. Taken together, these results suggest that NV-color-center defects may scatter phonons more strongly than models treating Nitrogen and vacancy as independent point defects, providing new insights into color-center-induced phonon-scattering mechanisms.

## Introduction

Diamond exhibits an exceptionally high thermal conductivity due to the low atomic mass of carbon, highly rigid C-C covalent bonds and a simple crystal structure.[1–3] Natural diamond yields a room-temperature thermal conductivity of approximately 2200 W $m^{-1}$ $K^{-1}$,[4–8] whereas reducing the abundance of $^{13}C$ isotopes suppresses point-defect scattering and can increase this value beyond 3000 W $m^{-1}$ $K^{-1}$.[5,9–15]

Therefore, isotope engineering has emerged as a powerful route for minimizing phonon scattering and maximizing thermal conductivity. This strategy has been successfully demonstrated not only in diamond but also in other high thermal conductivity materials. For instance, previous studies reported room temperature thermal conductivity exceeding 1600 W $m^{-1}$ $K^{-1}$ in isotopically enriched c-$^{10}BN$ and c-$^{11}BN$ (an ~82% enhancement), 600 W $m^{-1}$ $K^{-1}$ in $^{10}BP$ (an ~22% enhancement) and 1500 W $m^{-1}$ $K^{-1}$ in c-$^{11}BAs$ (an ~11% enhancement).[5,16,17] These results establish isotope scattering as an important resistance to heat transport, while exposing a follow-up question: once isotope scattering has been reduced, how significantly residual defects prevent a crystal from reaching its intrinsic thermal conductivity?

Color centers make this question particularly important in diamond. Diamond is a premier host for optically active defects, especially Nitrogen-Vacancy (NV) centers. Their coupled spin and optical states provide solid-state resources for quantum applications and single-photon generation.[18,19] These functional properties arise from impurity-vacancy complexes rather than isolated impurity atoms. The same local structure that enables quantum functionality may therefore introduce phonon scattering not represented by a model for independent point defects.

Previous calculations show why the distinction matters. Using an *ab initio* Green's function approach, nitrogen impurities, and especially vacancies have been found to suppress the thermal conductivity of diamond much more strongly than predicted by conventional Born approximation because defects also change local harmonic force constants.[20] Substitution of B in 3C-SiC was then predicted to breaks the local symmetry and produces resonant scattering roughly two orders of magnitude stronger than N substitution, despite the similar B-C and N-C mass difference.[21] The predicted strong scattering was subsequently observed experimentally in B-doped 3C-SiC with a nearly 50% reduction in the thermal conductivity at a boron concentration of $4\times10^{19}$ $cm^{-3}$.[22] These studies show that impurity concentration and mass difference alone are sometimes not sufficient to predict the strength of defect phonon scattering. This raises the possibility that the same issue may arise for complex defects in isotope-purified diamond.

Previous experimental evidence for nitrogen-related defects in diamond showed that the thermal conductivity of a lightly N-doped pink CVD diamond was strongly suppressed below 120 K, showing a maximum reduction by a factor of ~2.7 near 40 K.[2,12] Whether color-center-related defects can also strongly suppress room-temperature thermal conductivity in highly isotope-purified diamond remains unclear.

Here, we address this question by comparing two isotope-purified single crystal diamonds that have comparable crystalline quality but sharply different color center concentrations. Their thermal conductivities are characterized by time-domain thermoreflectance (TDTR), while photoluminescence (PL) and secondary-ion mass spectrometry (SIMS) are used to characterize their defect-related optical signatures and impurity compositions. Then a classical Callaway framework incorporating measured isotope and impurity concentrations was adopted to estimate this reduction in thermal conductivity.

**Results**

The two diamond samples we investigated are commercially available from Element Six and 3 mm by 3 mm in size, both denoted as B1 and B14 (Fig. 1a). We first tested whether their difference in thermal conductivity could originate from different crystalline quality. Both samples show sharp, symmetric (004) and (113) rocking curves (Figs. 1c and 1d); the (113) full width at half maximum (FWHM) values are 62.69 arcsec for B1 and 62.96 arcsec for B14.[23] Their first-order diamond Raman peaks are likewise narrow, with FWHM of approximately 2.6 cm$^{-1}$, further indicating high crystalline quality and little extended structural disorder.[23,24] The slight difference in peak positions are observed (1332.4 cm$^{-1}$ for B1 and 1332.7 cm$^{-1}$ for B14), which may be related to a difference in local residual strain and measurement errors. Overall, the nearly identical rocking curve width and Raman linewidth provide no evidence for a meaningful difference in crystalline quality and extended disorder. They do not exclude point defects or defect complexes, which can remain inconspicuous in these measurements while scattering heat-carrying phonons.

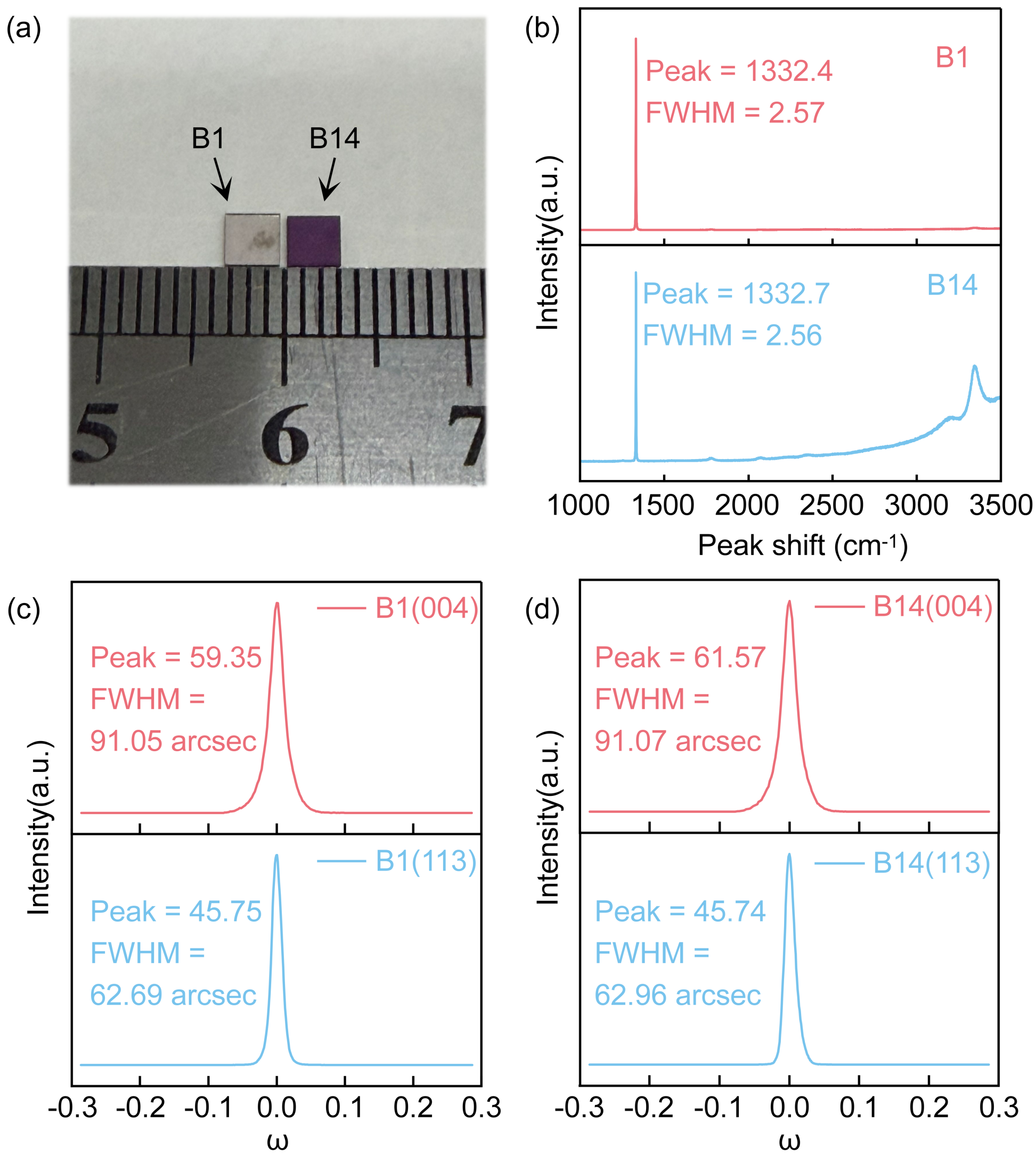

Figure 1. Structural and optical characterizations of B1 and B14. a) Photograph of the samples. b) Raman spectra. c, d) HRXRD rocking curves for the (004) and (113) reflections, respectively.

In contrast, PL spectroscopy exposes a clear difference in the defect-related optical signatures between the samples. Under 405 nm and 532 nm excitation, as shown in Figs. 2a and 2b, both samples show zero-phonon lines (ZPLs) associated with $NV^0$ (~575 nm) and $NV^-$ (~638 nm) centers.[25] Relative to its diamond Raman signal, under 532 nm excitation, the first order Raman line near 573 nm remains a visible shoulder in B1 but is obscured by the stronger $NV^0$ emission in B14, indicating a much stronger defect-related emission background in B14 (Fig. S1). B14 also exhibits a broad band near 467 nm, consistent with the broad feature near 3345 $cm^{-1}$ in its Raman spectrum and with vacancy-related defects.[26] These spectra show that the N-related defects in B14 is not totally described as isolated substitutional N and at least part of it is incorporated into optically active complexes.[27,28] In summary, B1 and B14 have comparable crystalline quality but distinctly different defect-related optical signatures.

To further compare the relative NV-related optical signals, the PL spectra measured under 405 nm excitation, whose diamond Raman peak is relatively obvious, are baseline-corrected, and the peak heights of the $NV^0$ ZPL and the diamond Raman signal were extracted, as shown in Figs. 2c and 2d. The relative NV concentrations can be estimated from fluorescence intensity ratio of renormalized NV ZPLs and diamond Raman peak signal, and an approximately linear relationship between this ratio and NV concentration can be assumed within a not-too-high concentration range.[29,30] In this framework, we can get:

$$R_i = \frac{H_{\mathrm{NV}^0,i}}{H_{\mathrm{Raman},i}} \tag{1}$$

$$F_{\mathrm{NV}^0}^{\mathrm{opt}} = \frac{R_{\mathrm{B14}}}{R_{\mathrm{B1}}} \approx \frac{n_{\mathrm{NV}^0,\mathrm{B14}}}{n_{\mathrm{NV}^0,\mathrm{B1}}} \tag{2}$$

Here, $H_{\mathrm{NV}^0,i}$ and $H_{\mathrm{Raman},i}$ denote the baseline-corrected peak heights of the $NV^0$ ZPL and the diamond Raman peak, respectively, $R_i$ is the corresponding Raman-normalized peak-height ratio, and $n_{\mathrm{NV}^0,i}$ represents the $NV^0$ number concentration. The extracted ratios are $R_{\mathrm{B1}} \approx 1.28$ and $R_{\mathrm{B14}} \approx 9.11$, yielding a relative optical-signal enhancement factor of $F_{\mathrm{NV}^0}^{\mathrm{opt}} \approx 7.12$, this ratio provides a semiquantitative estimate of the relative $NV^0$ concentration. For the subsequent analysis, we further assume that the fraction of $NV^0$ and $NV^-$ centers is approximately the same in B1 and B14. Under this assumption, the ratio of the total NV concentrations can be approximated by the $NV^0$ concentration ratio, yielding an estimated approximately 7.1-fold higher total NV concentration in B14 than in B1.

SIMS further quantifies the pronounced difference in the impurity composition of B1 and B14 (Fig. 2e). B1 contains approximately $5.5\times10^{17}$ atoms cm$^{-3}$ $^{13}$C, corresponding to ~0.00031%. B14 remains isotope enriched but contains ~$6.3\times10^{20}$ atoms cm$^{-3}$ $^{13}$C (~0.36%), markedly higher than B1. The N signal in B1 are close to their instrumental background levels, whereas B14 contains ~$7.1\times10^{17}$ atoms cm$^{-3}$ N. These impurity trends are consistent with the stronger color center-related PL spectra observed in B14. Other potential impurities, including hydrogen, boron, silicon, and oxygen, remain comparatively low or their contributions are not expected to dominate the following thermal analysis, which are shown in Fig. S2. These measured $^{13}$C and N concentrations provide the composition-dependent inputs for the subsequent Callaway analysis.

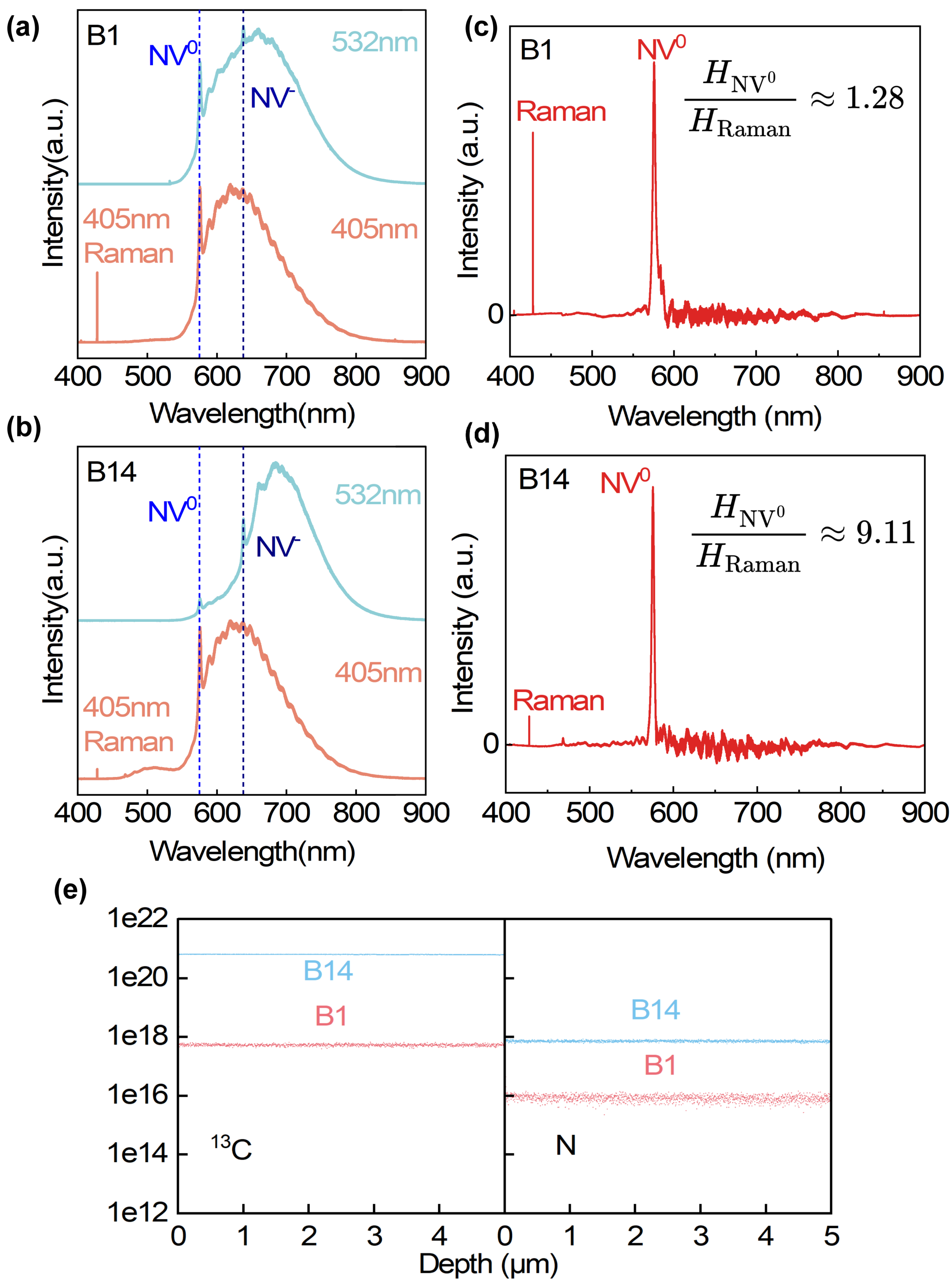

Figure 2. PL and SIMS resuts of B1 and B14. a, b) PL spectra measured under 405 nm and 532 nm excitation, showing the zero-phonon lines of $NV^0$ (~575 nm) and $NV^-$ (~638 nm). Under 532 nm excitation, the diamond Raman peak near 573 nm is largely obscured by the strong $NV^0$ emission in B14; an expanded view is provided in Fig. S1.

c, d) Baseline-corrected PL spectra of B1 and B14, under 405 nm excitation, respectively, used to extract the peak heights of the $NV^0$ and diamond Raman signal. The corresponding peak-height ratios, $H_{NV^0}/H_{Raman}$, are approximately 1.28 for B1 and 9.11 for B14, giving a relative enhancement factor of approximately 7.12, providing a semiquantitative measure of relative $NV^0$ concentration. e) SIMS depth profiles for $^{13}C$ and N in B1 and B14.

Multi-frequency and multi-spot-size TDTR measurements were performed on both samples to examine possible through-plane and lateral quasi-ballistic effects (Figs. 3a and 3b). The TDTR method is described in Refs.[30,31] Both two-tint and two-color TDTR configuration were used in the present work. For multi-frequency measurements, the modulation frequency sets the through plane thermal length scale, commonly described in the diffusive limit by the thermal penetration depth, $d_{p,z} = \left(\kappa_z / \pi C f_{\mathrm{mod}}\right)^{1/2}$, where $C$ is the volumetric heat capacity.[32] Owing to the high thermal diffusivity of diamond, $d_{p,z}$ remains several micrometers across the investigated modulation frequencies. Within experimental uncertainty, the apparent thermal conductivities of B1 and B14 show no systematic dependence on $f_{\mathrm{mod}}$, indicating that through-plane quasi-ballistic effects are small under these conditions.[33,34]

For multi-spot-size measurements, the laser spot size sets a lateral thermal length scale. First-principles calculations place the room-temperature phonon mean free path spectrum of diamond at approximately 0.3-25 μm, with phonons above 5 μm contributing less than 10% of the total thermal conductivity.[8] B1 shows only a modest decrease in the apparent κ when the spot size is reduced, consistent with a limited quasi-ballistic contribution over the present stage.[34] B14 shows an even weaker spot-size dependence, suggesting that these long-MFP contributions may have already been truncated by internal scattering. This observation does not identify the scatter source by itself, but it is consistent with the denser defect revealed by PL and SIMS.

We further examined whether the Al–diamond interface could contribute to the difference in the extracted thermal conductivity. As shown in Figure 3b, for the original Al transducer, the thermal boundary conductance (TBC) values of B1 and B14 are approximately 100-120 MW $m^{-2}$ $K^{-1}$ and show no systematic dependence on laser spot size. According to Fig. S2, the TBC shows frequency-independence. The similar TBC values of the two samples also indicate that their large difference in thermal conductivity cannot be attributed to differences in the Al-diamond interface. To further test the robustness of the measurements, the Al transducers were removed using a NaOH solution and redeposited after the first round of measurements. The thermal conductivities subsequently measured using a 9.0 μm spot radius agree with the large spot results obtained from the first deposition within experimental uncertainty, demonstrating good reproducibility of the extracted diamond thermal conductivity. The TBC values after redeposition are somewhat lower, approximately 80-95 MW $m^{-2}$ $K^{-1}$, likely reflecting variations in interface preparation between the two metal coating processes. This modest change does not affect the main conclusions because the corresponding TBC was explicitly accounted for in each TDTR fit, while the extracted thermal conductivities remained essentially unchanged.

Accounting for these measurement and interface effects, we next compare the intrinsic thermal conductivities of B1 and B14. Results show B1 reaches ~3200 W $m^{-1}$ $K^{-1}$ at room temperature, as expected for the benefit of ultrahigh $^{12}C$ enrichment and low impurity concentrations. By contrast, B14 reaches only ~2100 W $m^{-1}$ $K^{-1}$, even though its $^{13}C$ fraction of 0.36% is well below the natural abundance of ~1.1%.

To test whether conventional defect phonon scattering mechanism can explain this reduction, we analyzed the temperature-dependent TDTR data with the classical Callaway model.[35–38] The model includes Normal (N) and Umklapp (U) three phonon scattering processes, boundary scattering, and point-defect scattering. The full expressions are provided in Supplementary Information.

An intrinsic baseline for phonon-phonon scatterings and point-defect scatterings was first obtained by jointly fitting the high-temperature thermal-conductivity data for a natural diamond (Nat) and B1. Because the two samples share almost the same diamond lattice, the N process coefficient $A_N$ and the scaling parameter $T_D/\alpha$ were shared between the two samples. The Umklapp scattering $A_U$, the point-defect scattering coefficient $A_{pd}$, and the effective surface roughness parameter $\eta$ were fitted separately. The boundary scattering length $l_b$ was fixed to the physical thickness of each sample.

An example of all fitting results including natural diamond, B1 and B14 is demonstrated in Fig. S6.

Notably, B14 was then evaluated without refitting the intrinsic terms. We retained the lattice scattering parameters obtained for B1 and recalculated the point-defect coefficient $A_{pd}$ taking the $^{13}C$ and N concentrations measured by SIMS and using the classical expression of Turk and Klemens,[2,39]

$$A_{pd} = \sum_i c_i \left( \Delta M_i / M + 2\gamma\beta_i \right)^2 \frac{k_B{}^4 V_0}{4\pi v^3 \hbar^4} \quad (3)$$

where $c_i$ is the atomic percentage of impurity species $i$ obtained from SIMS, $\Delta M_i / M$ is the relative mass difference between impurity species $i$ and the average host lattice. $\gamma$ is the Grüneisen parameter corresponding to crystal lattice, which is 1.1 for diamond[40]. $\beta$ is the relative difference in atomic volumes between impurity atom and carbon in the diamond lattice, accounting for the additional strain field contribution associated with local lattice distortion, which is 0.35 for Nitrogen[12,41].

Although the vacancy concentration in B14 cannot be directly quantified from the present SIMS and PL measurements, vacancies are known to be strong phonon scatterers in diamond[12,20,42]. To examine their possible contribution to the reduced thermal conductivity of B14, we compared four vacancy-scattering models within the Callaway framework, denoted as Model I-IV. Before calculations, the vacancy concentration was initially set to be equal to the total nitrogen concentration, assuming that each nitrogen atom forms an NV color center with one associated vacancy. This represents an upper-bound value for the NV color center concentration rather than experimentally established, benefiting for qualitative interpretation of subsequent calculated results.

In Model I, initially, only the mass perturbation was considered, while the strain-field contribution was neglected. The corresponding effective perturbation is

$$S_V = \frac{\Delta M_V}{M} = -1 \tag{4}$$

giving a vacancy-related disorder term. Even with the upper-bound assumption $c_V = c_N$, the predicted room-temperature thermal conductivity of B14 remains ~2529 W m$^{-1}$ K$^{-1}$, substantially higher than the measured value of ~ 2100 W m$^{-1}$ K$^{-1}$. This result suggests that the mass perturbation alone is insufficient to account for the observed suppression, and strain-field contribution is significant.

In Model II, we additionally considered the local lattice distortion associated with a vacancy. Under the conventional volume mismatch definition, assuming an effective vacancy volume of zero gives $\beta_V = -1$. The perturbation term in Eq. (3) becomes:

$$S_V = \frac{\Delta M_V}{M} + 2\gamma\beta_V = -1 + 2(1.1)(-1) = -3.2 \tag{5}$$

and the corresponding vacancy-related disorder term is $\mathrm{c_V S_V^2} = 10.24\mathrm{c_V}$. With $c_V = c_N$, this model predicts a room-temperature thermal conductivity of ~ 2130 W m$^{-1}$ K$^{-1}$, close to the experimental value. However, this agreement relies on the assumption that every nitrogen atom forms a NV color center. With actual NV concentration, the lower vacancy proportion would weaken the predicted scattering and increase the calculated thermal conductivity. Thus, the previous agreement at room temperature does not establish the validity of this concentration assumption.

To examine a stronger vacancy-scattering description, Model III adopted the first-principles Green's-function and T-matrix vacancy-scattering result[20]. In this parameterization, the effective perturbation replaces the entire mass-and-strain term in Eq. (3), giving

$$S_{V,\mathrm{eff}} \equiv \frac{\Delta M_{\mathrm{eff}}}{M} = -6 \quad (6)$$

And thus $c_V S_{V,\mathrm{eff}}^2 = 36 c_V$. With $c_V = c_N$, the predicted room-temperature thermal conductivity decreases to ~1593 W m$^{-1}$ K$^{-1}$, markedly below the measured value. This result poses a challenge to the concentration assumption, indicating that the 100% vacancy concentration produces excessive suppression in the predicted thermal conductivity.

In Model IV, we retained the effective perturbation $\Delta M_{\mathrm{eff}}/M = -6$, but treated the vacancy concentration as an adjustable parameter. Using the measured room-temperature thermal conductivity of B14 as the calibration target, we obtained $c_V \approx 30\% c_N$, corresponding to a point-defect scattering coefficient $A_{pd}$ of approximately 3.88*10$^{-3}$ s$^{-1}$ K$^{-4}$. The resulting room-temperature thermal conductivity is ~2107 W m$^{-1}$ K$^{-1}$, in close agreement with experiment.

However, all four models above fail to reproduce well the relationship between temperature and B14 thermal conductivity obtained from experimental results. The four models' B14 $A_{pd}$ and their corresponding room-temperature predictions are summarized in Fig. 3(c), while Fig. 3(d) compares their calculated temperature-dependent thermal conductivities with the experimental data for B14. Although Models II and IV reproduce the room-temperature value reasonably well, none of the four models satisfactorily captures the measured temperature dependence over the full experimental range. This suggests that the Callaway model adopted above maybe limited to describe the interaction between phonons and color centers.

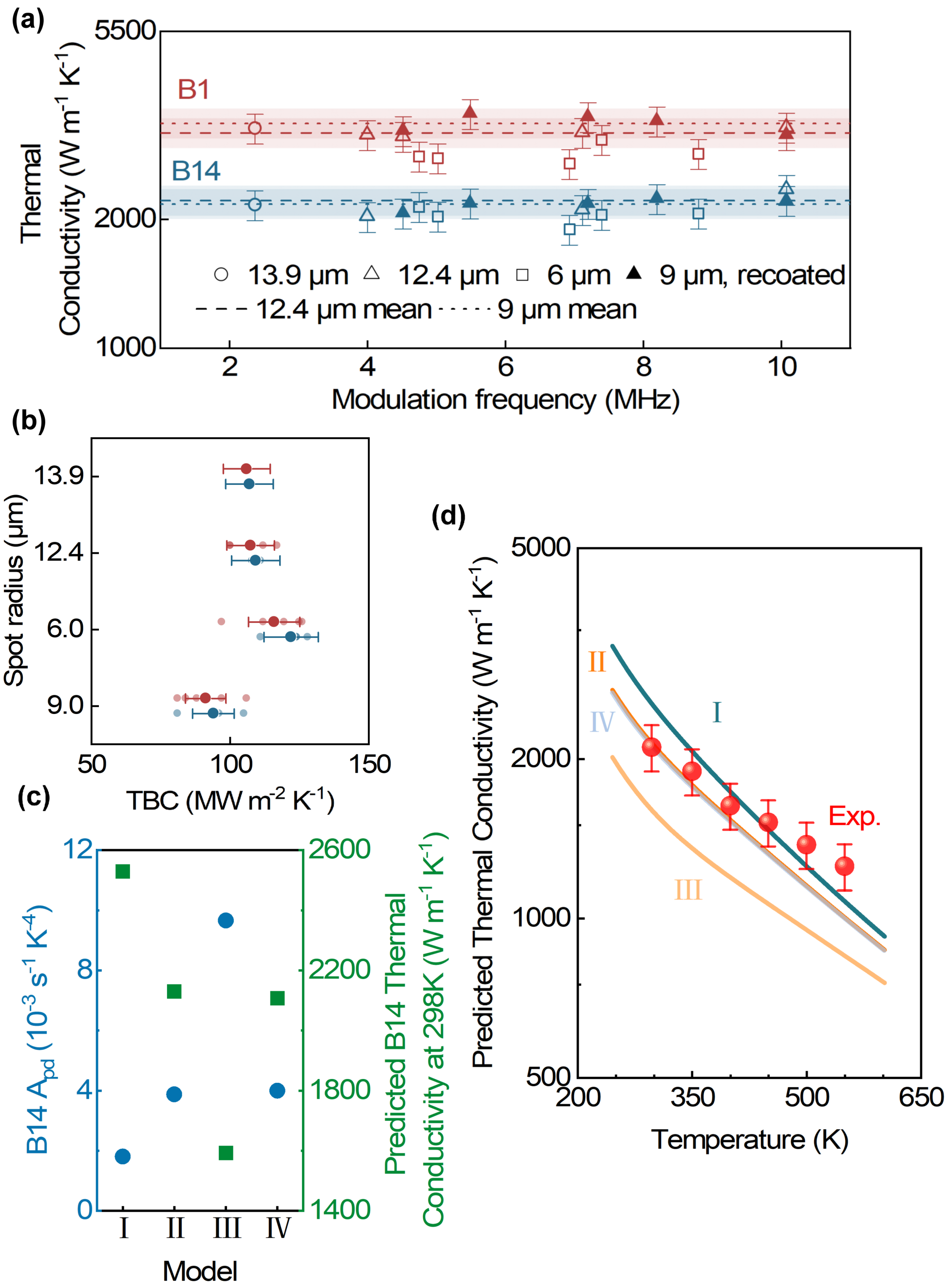

Figure 3. TDTR measurements and Callaway analysis. a) Room-temperature thermal conductivity and b) Al-diamond TBC of B1 (red) and B14 (blue) measured at different modulation frequencies and laser spot sizes. Filled triangles denote measurements with a 9.0 μm spot radius after the Al transducer was removed and redeposited. c) Calculated point-defect scattering coefficient $A_{pd}$ and predicted thermal conductivity of B14 at 298

K for four vacancy-scattering models. (d) Measured temperature-dependent thermal conductivity of B14 compared with the predictions of models I–IV.

To investigate this limitation and assess the plausibility of the concentration required by Model IV, we compared the PL and SIMS results discussed at Fig. 2. As described above, a semiquantitative estimate form PL signals indicates that the total NV concentration in B14 is ~7.12 times that in B1. In contrast, the total nitrogen concentration in B14 is approximately 92 times that in B1, based on the nominal SIMS values. Based on these two values and 30% assumption from Model IV, we can calculate the approximate NV proportion in B1 to be

$$f_{\mathrm{B1}}^{\mathrm{implied}} = f_{\mathrm{B14}}^{\mathrm{model}} \frac{n_{\mathrm{N,B14}}/n_{\mathrm{N,B1}}}{n_{\mathrm{NV,tot,B14}}/n_{\mathrm{NV,tot,B1}}} = \frac{30\% \times 92}{7.12} \approx 3.88 > 1 \tag{7}$$

Here; $f_{\mathrm{B14}}^{\mathrm{model}} = 30\%$is the fraction assumed in Model IV, $n_{\mathrm{N,B14}}/n_{\mathrm{N,B1}} \approx 92$ is the total nitrogen concentration ratio and $n_{\mathrm{NV,tot,B14}}/n_{\mathrm{NV,tot,B1}} \approx 7.12$ is the semiquantitatively estimated total NV concentration ratio based on PL. Since $f_{\mathrm{B1}} \leq 1$, the fraction from Model IV in B14 is greatly overestimated. This discrepancy would be even greater if, as is likely, only a fraction of the nitrogen in B1 forms NV color centers. If these NV color centers in B14 produce a total scattering contribution comparable to that represented by Model IV, their scattering strength for single color center must be greater than that captured by the model. This suggests that the observed thermal conductivity reduction cannot be explained simply by adding the scattering contributions of isolated nitrogen and vacancy point defects.

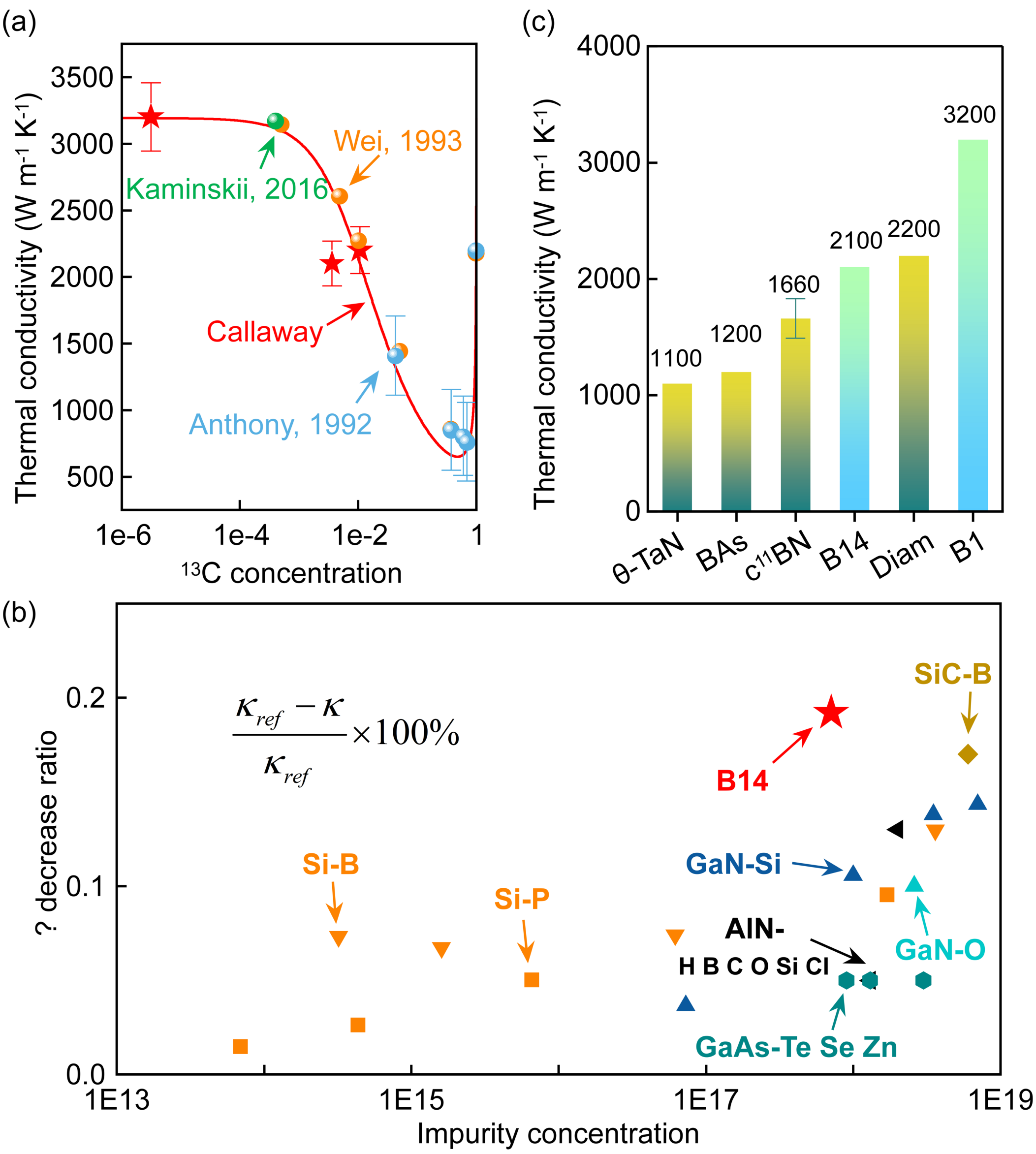

Figure 4. a) Room temperature thermal conductivity versus $^{13}$C concentration. The Callaway curve from this work is compared with data from Anthony,[43] Wei,[9] and Kaminskii[44]. b) Relative thermal conductivity reduction in different semiconductors plotted against impurity concentration;[22,40,45–49] sample B14 is plotted using the nitrogen concentration. c) Room-temperature thermal conductivity of representative high thermal conductivity bulk materials, including θ-TaN,[50] BAs,[3] c$^{11}$BN,[5] B14, diamond, and B1.

To disentangle the effect of isotope composition, Fig. 4a shows a series of experimental room-temperature single-crystal diamond thermal conductivities and the calculated curve varying the $^{13}C$ abundance while setting the N concentration to zero. Experimental and calculated results are in great agreement, but B14 falls far below the value expected from its $^{13}C$ fraction, further confirm the strong scattering effect from color centers. These results suggest that conventional mass and strain disorder models may not fully capture the phonon scattering associated with NV color center defects. Such complexes may perturb local bonding and interatomic force constants across multiple lattice sites. However, the present measurements do not identify a specific defect structure or establish a particular scattering mechanism,

Fig. 4b places the additional thermal conductivity reduction observed in B14 in a broader materials context by comparing the relative thermal conductivity reduction caused by impurities in different semiconductors. At a total N concentration of ~ $7.1 \times 10^{17}$ $cm^{-3}$, B14 shows an additional κ decrease of more than 19%. Comparable decreases in the literature included in this comparation generally occur at impurity concentrations near $10^{19}$-$10^{21}$ $cm^{-3}$, whereas Si: B, Si: P, and GaN: Si at similar or higher concentrations show smaller reductions. The horizontal coordinate for B14 is conservative because it uses the total SIMS N concentration. Only a fraction of those N atoms would belong to optically active complexes, so the actual complex concentration cannot be inferred and may be lower. Because the plotted materials differ in host crystal, dopant, and reference conductivity, the comparison is qualitative rather than a universal scattering cross section. Even with that limitation, B14 occupies an unusual region of the plot, exhibiting a substantial conductivity loss at a comparatively low measured impurity concentration.

Fig. 4c compares B1 and B14 with other representative ultrahigh-thermal-conductivity materials. B1 reaches about 3200 W $m^{-1}$ $K^{-1}$, substantially higher than natural diamond and other leading thermal-management materials such as BAs and isotope-enriched cBN. In contrast, B14, despite being isotope-purified, falls close to natural diamond attributed to its strong defect-induced thermal degradation. These results highlight that the thermal performance of isotope-enriched diamond depends not only on isotopic purification but also on residual defect populations, underscoring the importance of defect control when optimizing diamond for heat transport.

**Conclusions**

In summary, we experimentally demonstrate substantial thermal conductivity suppression in isotope-purified diamond with different color center concentrations. Structural and compositional characterizations show B1 and B14 have nearly identical HR-XRD rocking-curve widths and Raman linewidths but very different impurity and color-center signatures. B1 reaches a room-temperature thermal conductivity of approximately 3200 W $m^{-1}$ $K^{-1}$, while B14 reaches only approximately 2100 W $m^{-1}$ $K^{-1}$ despite a $^{13}C$ abundance of approximately 0.36%. Four models based on Callaway calculation are adopted. Although Model IV approaches the measured room-temperature value under its assumptions, it fails to capture the full temperature dependence; moreover, its assumed defect concentration is difficult to reconcile with the PL and SIMS data. Taken together, these results suggest that color center may introduce stronger phonon scattering than predicted by conventional isolated-point-defect models. These findings provide insights into new mechanism related to color centers in phonon transport and highlight the need to eliminate color center impurities when engineering isotope-purified diamond for high thermal conductivity.

## Conflict of Interest

The authors declare that they have no conflict of interest.

## Acknowledgements

This work was supported by the National Key Research and Development Program of China (2024YFA1207901) and the National Natural Science Foundation of China (62574007, T2550270).

## Data availability

The data that support the findings of this study are available from the corresponding authors upon reasonable request.

**References**

[1] G.A. Slack, "Nonmetallic crystals with high thermal conductivity," J. Phys. Chem. Solids **34**, 321–335 (1973).

[2] A.V. Inyushkin, A.N. Taldenkov, V.G. Ralchenko, G. Shu, B. Dai, A.P. Bolshakov, A.A. Khomich, E.E. Ashkinazi, K.N. Boldyrev, A.V. Khomich, J. Han, V.I. Konov, and J. Zhu, "Thermal conductivity of pink CVD diamond: Influence of nitrogen-related centers," J. Appl. Phys. **133**(2), 025102 (2023).

[3] J.S. Kang, M. Li, H. Wu, H. Nguyen, and Y. Hu, "Experimental observation of high thermal conductivity in boron arsenide," Science **361**(6402), 575–578 (2018).

[4] Z. Cheng, Z. Huang, J. Sun, J. Wang, T. Feng, K. Ohnishi, J. Liang, H. Amano, and R. Huang, "(Ultra)wide bandgap semiconductor heterostructures for electronics cooling," Appl. Phys. Rev. **11**(4), 041324 (2024).

[5] K. Chen, B. Song, N.K. Ravichandran, Q. Zheng, X. Chen, H. Lee, H. Sun, S. Li, G.A.G. Udalamatta Gamage, F. Tian, Z. Ding, Q. Song, A. Rai, H. Wu, P. Koirala, A.J. Schmidt, K. Watanabe, B. Lv, Z. Ren, L. Shi, D.G. Cahill, T. Taniguchi, D. Broido, and G. Chen, "Ultrahigh thermal conductivity in isotope-enriched cubic boron nitride," Science **367**(6477), 555–559 (2020).

[6] Z. Cheng, T. Bougher, T. Bai, S.Y. Wang, C. Li, L. Yates, B.M. Foley, M. Goorsky, B.A. Cola, F. Faili, and S. Graham, "Probing growth-induced anisotropic thermal transport in high-quality CVD diamond membranes by multifrequency and multiple-spot-size time-domain thermoreflectance," ACS Appl. Mater. Interfaces **10**(5), 4808–4815 (2018).

[7] Z. Cheng, J. Liang, K. Kawamura, H. Zhou, H. Asamura, H. Uratani, J. Tiwari, S. Graham, Y. Ohno, Y. Nagai, T. Feng, N. Shigekawa, and D.G. Cahill, "High thermal conductivity in wafer-scale cubic silicon carbide crystals," Nat. Commun. **13**(1), 7201 (2022).

[8] H. Zhou, J. Tiwari, Q. Xu, M.F.R.J. Islam, T. Abedien, and T. Feng, "Effective phonon mean free paths for thermal modeling of modern semiconductors," IEEE Trans. Electron Devices **73**(4), 2430–2439 (2026).

[9] L. Wei, P.K. Kuo, R.L. Thomas, T.R. Anthony, and W.F. Banholzer, "Thermal conductivity of isotopically modified single crystal diamond," Phys. Rev. Lett. **70**(24), 3764–3767 (1993).

[10] J.R. Olson, R.O. Pohl, J.W. Vandersande, A. Zoltan, T.R. Anthony, and W.F. Banholzer, "Thermal conductivity of diamond between 170 and 1200 K and the isotope effect," Phys. Rev. B **47**(22), 14850–14856 (1993).

[11] L. Lindsay, D.A. Broido, and T.L. Reinecke, "Phonon-isotope scattering and thermal conductivity in materials with a large isotope effect: A first-principles study," Phys. Rev. B **88**(14), 144306 (2013).

[12] A.V. Inyushkin, A.N. Taldenkov, V.G. Ralchenko, A.P. Bolshakov, and A.V. Khomich, "Isotope Effect in Thermal Conductivity of Polycrystalline CVD-Diamond: Experiment and Theory," Crystals **11**(4), 322 (2021).

[13] H. Zhou, S. Zhou, Z. Hua, K. Bawane, and T. Feng, "Impact of classical statistics on thermal conductivity predictions of BAs and diamond using machine learning molecular dynamics," Appl. Phys. Lett. **125**(17), 172202 (2024).

[14] A.V. Inyushkin, V.G. Ralchenko, A.P. Bolshakov, A.A. Khomich, D.A. Chernodubov, A.N. Taldenkov, V.V. Saraykin, and S.Ya. Kilin, "Thermal conductivity of chemical vapor deposition diamond enriched with 13C isotope," J. Appl. Phys. **137**(12), 125109 (2025).

[15] N.V. Novikov, A.P. Podoba, S.V. Shmegera, A. Witek, A.M. Zaitsev, A.B. Denisenko, W.R. Fahrner, and M. Werner, "Influence of isotopic content on diamond thermal conductivity," Diam. Relat. Mater. **8**(8–9), 1602–1606 (1999).

[16] J. Kim, D. Lee, H. Wu, and J.S. Kang, "Isotope-Enriched Cubic Boron Arsenide with Ultrahigh Thermal Conductivity," Adv. Sci. **12**(25), 2502544 (2025).

[17] Y. Zhu, A.L.N. Kondusamy, K. Chen, et al., "Vapor-Flux Growth of c-BP Single Crystals With Concurrently High Electrical Resistivity and Isotope-Enhanced High Thermal Conductivity," Advanced Materials **38**(12), e17389 (2026).

[18] D.D. Awschalom, R. Hanson, J. Wrachtrup, and B.B. Zhou, "Quantum technologies with optically interfaced solid-state spins," Nat. Photonics **12**(9), 516–527 (2018).

[19] C. Bradac, W. Gao, J. Forneris, M.E. Trusheim, and I. Aharonovich, "Quantum nanophotonics with group IV defects in diamond," Nat. Commun. **10**(1), 5625 (2019).

[20] N.A. Katcho, J. Carrete, W. Li, and N. Mingo, "Effect of nitrogen and vacancy defects on the thermal conductivity of diamond: An *ab initio* Green's function approach," Phys. Rev. B **90**(9), 094117 (2014).

[21] A. Katre, J. Carrete, B. Dongre, G.K.H. Madsen, and N. Mingo, "Exceptionally Strong Phonon Scattering by B Substitution in Cubic SiC," Phys. Rev. Lett. **119**(7), 075902 (2017).

[22] Z. Huang, J. Liang, Y. Wang, Z. Sun, N. Shigekawa, M. Li, R. Wang, and Z. Cheng, "Experimental observation of extremely strong defect–phonon scatterings in cubic SiC single crystals," Appl. Phys. Rev. **13**(1), 011431 (2026).

[23] T. Zhu, W. Zhang, Y. Zhu, C. Liu, P. Sun, G. Yang, Q. Yuan, K. Yang, Y. Lu, H. Song, J. Yi, C.-T. Lin, W. Lian, K. Nishimura, Y. Wang, and N. Jiang, "Fabrication, microstructure and optical properties of a 30 × 30 × 1 mm3-sized mosaic single crystal diamond for IR window," Diam. Relat. Mater. **160**, 113067 (2025).

[24] S. Prawer, and R.J. Nemanich, "Raman spectroscopy of diamond and doped diamond," Philos. Trans. R. Soc. Lond. Ser. Math. Phys. Eng. Sci. **362**(1824), 2537–2565 (2004).

[25] B. Dong, C. Shi, Z. Xu, K. Wang, H. Luo, F. Sun, P. Wang, E. Wu, K. Zhang, J. Liu, Y. Song, and Y. Fan, "Temperature dependence of optical centers in Ib diamond characterized by photoluminescence spectra," Diam. Relat. Mater. **116**, 108389 (2021).

[26] A.M. Zaitsev, N.M. Kazuchits, K.S. Moe, J.E. Butler, O.V. Korolik, M.S. Rusetsky, and V.N. Kazuchits, "Luminescence of brown CVD diamond: 468 nm luminescence center," Diam. Relat. Mater. **113**, 108255 (2021).

[27] M.N.R. Ashfold, J.P. Goss, B.L. Green, P.W. May, M.E. Newton, and C.V. Peaker, "Nitrogen in Diamond," Chem. Rev. **120**(12), 5745–5794 (2020).

[28] Y. Zheng, C. Li, J. Liu, J. Wei, and H. Ye, "Diamond with nitrogen: states, control, and applications," Funct. Diam. **1**(1), 63–82 (2021).

[29]A. Jarmola, A. Berzins, J. Smits, K. Smits, J. Prikulis, F. Gahbauer, R. Ferber, D. Erts, M. Auzinsh, and D. Budker, "Longitudinal spin-relaxation in nitrogen-vacancy centers in electron irradiated diamond," Appl. Phys. Lett. 107(24), 242403 (2015).

[30] M. Mrózek, A. M. Wojciechowski, and W. Gawlik, "Characterization of strong NV− gradient in the e-beam irradiated diamond sample," Diamond Relat. Mater. 120, 108689 (2021).

[31] J. Liu, Z. Huang, L. Yang, Y. Zhang, X. Zhang, K. Zhang, X. Guo, Y. Wang, H. Zhou, J. Zhang, W. Wang, Y. Hao, and Z. Cheng, "Ion Implantation Enhanced Nucleation Facilitates Heat Transport across Atomically-Sharp Semiconductor Interfaces," (2026).

[32] P. Jiang, X. Qian, and R. Yang, "Tutorial: time-domain thermoreflectance (TDTR) for thermal property characterization of bulk and thin film materials," J. Appl. Phys. **124**(16), 161103 (2018).

[33] R.B. Wilson, and D.G. Cahill, "Anisotropic failure of Fourier theory in time-domain thermoreflectance experiments," Nat. Commun. **5**(1), 5075 (2014).

[34] R.B. Wilson, and D.G. Cahill, "Limits to Fourier theory in high thermal conductivity single crystals," Appl. Phys. Lett. **107**(20), 203112 (2015).

[35] A.V. Inyushkin, A.N. Taldenkov, V.G. Ralchenko, A.P. Bolshakov, A.V. Koliadin, and A.N. Katrusha, "Thermal conductivity of high purity synthetic single crystal diamonds," Phys. Rev. B **97**(14), 144305 (2018).

[36] J. Callaway, "Model for lattice thermal conductivity at low temperatures," Phys. Rev. **113**(4), 1046–1051 (1959).

[37] Z. Cheng, B.M. Foley, T. Bougher, L. Yates, B.A. Cola, and S. Graham, "Thermal rectification in thin films driven by gradient grain microstructure," J. Appl. Phys. **123**(9), 095114 (2018).

[38] Z. Cheng, Y.R. Koh, A. Mamun, J. Shi, T. Bai, K. Huynh, L. Yates, Z. Liu, R. Li, E.

Lee, M.E. Liao, Y. Wang, H.M. Yu, M. Kushimoto, T. Luo, M.S. Goorsky, P.E. Hopkins, H. Amano, A. Khan, and S. Graham, "Experimental observation of high intrinsic thermal conductivity of AlN," Phys. Rev. Mater. **4**(4), 044602 (2020).

[39] L.A. Turk, and P.G. Klemens, "Phonon scattering by impurity platelet precipitates in diamond," Phys. Rev. B **9**(10), 4422–4428 (1974).

[40] R. Berman, P.R.W. Hudson, and M. Martinez, "Nitrogen in diamond: evidence from thermal conductivity," J. Phys. C Solid State Phys. **8**(21), L430–L434 (1975).

[41] A.R. Lang, and G. Pang, "On the Dilatation of Diamond by Nitrogen Impurity Aggregated in A Defects," (1998).

[42] C.A. Ratsifaritana and P.G. Klemens, "Scattering of phonons by vacancies," Int. J. Thermophys. 8(6), 737–750 (1987).

[43] T.R. Anthony, and W.F. Banholzer, "Properties of diamond with varying isotopic composition," Diam. Relat. Mater. **1**(5–6), 717–726 (1992).

[44] A.A. Kaminskii, V.G. Ral'chenko, H. Yoneda, A.P. Bol'shakov, and A.V. Inyushkin, "Stimulated Raman scattering-active isotopically pure $^{12}C$ and $^{13}C$ diamond crystals: A milestone in the development of diamond photonics," JETP Lett. **104**(5), 347–352 (2016).

[45] X. Fan, Z. Zhang, J. Zhu, K. Yuan, J. Zhou, X. Zhang, and D. Tang, "Systematic investigations on doping dependent thermal transport properties of single crystal silicon by time-domain thermoreflectance measurements," Int. J. Therm. Sci. **177**, 107558 (2022).

[46] P.P. Paskov, M. Slomski, J.H. Leach, J.F. Muth, and T. Paskova, "Effect of Si doping

on the thermal conductivity of bulk GaN at elevated temperatures – theory and experiment," AIP Adv. **7**(9), 095302 (2017).

[47] A. Jeżowski, O. Churiukova, J. Mucha, T. Suski, I.A. Obukhov, and B.A. Danilchenko, "Thermal conductivity of heavily doped bulk crystals GaN:O. Free carriers contribution," Mater. Res. Express **2**(8), 085902 (2015).

[48] R. Rounds, B. Sarkar, D. Alden, Q. Guo, A. Klump, C. Hartmann, T. Nagashima, R. Kirste, A. Franke, M. Bickermann, Y. Kumagai, Z. Sitar, and R. Collazo, "The influence of point defects on the thermal conductivity of AlN crystals," J. Appl. Phys. **123**(18), 185107 (2018).

[49] X. Qian, P. Jiang, and R. Yang, "Anisotropic thermal conductivity of 4H and 6H silicon carbide measured using time-domain thermoreflectance," Mater. Today Phys. **3**, 70–75 (2017).

[50] S. Li, C. Su, Z. Qin, A. Alatas, M. Kunz, T. Yamada, S.D. Kelly, M.H. Upton, A. Gironda, J. Zhao, B. Kalkan, W. Yang, T. Aoki, and Y. Hu, "Metallic θ-phase tantalum nitride has a thermal conductivity triple that of copper," Science **391**(6786), 707–711 (2026).